\documentclass[11pt,twoside]{article}
\usepackage{asp2010}

\def\h{\hbox{\kern 0.20em $^{\rm h}$}}
\def\m{\hbox{\kern 0.20em $^{\rm m}$}}
\def\s{\hbox{\kern 0.20em $^{\rm s}$}}
\def\arcdeg{\deg}

\resetcounters

\begin{document}

\title{Targeted deep surveys of high Galactic latitude HI with the GBT}
\author{K. Blagrave$^1$, Felix. J. Lockman$^2$, and P. G. Martin$^1$
\affil{$^1$Canadian Institute for Theoretical Astrophysics, University of Toronto, Toronto, Ontario, M5S 3H8, Canada}
\affil{$^2$National Radio Astronomy Observatory, P. O. Box 2, Green Bank, West Virginia 24944, USA}
}

\begin{abstract}
Over 800 sq. deg. of high Galactic latitude sky have been mapped at 21 cm with the Robert C. Byrd Green Bank Telescope (GBT). An improved knowledge of the telescope's beam characteristics has allowed us to reliably map not only regions of high column density, but also such regions as ELAIS N1, a targeted Spitzer field, which have very low HI column density. The additional fields we have observed cover a cross-section of dynamically and chemically interesting regions as indicated by the presence of intermediate/high velocity gas and/or anomalous far-IR (dust) colour.
\end{abstract}

\section{Introduction}
At high Galactic latitudes, the column density of atomic hydrogen is known to correlate very strongly with far infrared emission from dust  \citep{boulanger_1996}.  However, it has also been observed that HI clouds that have large peculiar velocities  (IVCs and HVCs) can have a significantly different FIR/HI ratio than normal disk gas \citep{wakker_1986, herbstmeier_1993, mamd_2005, peek_2009}. Deviations from the standard FIR/HI ratio are also  sometimes found in isolated ISM components.  
Progress in understanding of the dust at high Galactic latitudes, and its connection to processes in the interstellar medium, hinges on an understanding of the gas-dust correlations at the various observed infrared and sub-mm wavelengths.
With the Planck and Herschel satellites now returning sub-mm wavelength data at considerably higher angular resolution than the currently available all-sky HI surveys (the Leiden/Argentine/Bonn (LAB) HI survey  \citep{kalberla_2005}), it is vital to have comparable resolution, high sensitivity 21cm HI data available to understand the Galactic dust component of the maps.  

\section{Observations}

  We have undertaken a series of targeted observations of the Galactic HI component using the 100-m Robert C. Byrd Green Bank Telescope (GBT) at NRAO.
The GBT has an unblocked aperture with a high main beam efficiency and reduced far sidelobes compared with conventional telescopes.  Accurate calibration of the GBT data is discussed in \S~\ref{datareduction}.

Maps of Galactic HI in the 21cm line were made  on-the-fly at constant Galactic latitude, $b$, or constant declination, $\delta$, 
 on a $3\farcm5$ grid, slightly finer than Nyquist sampling for the $9\farcm1$  beam of the GBT.  
The 2.5 MHz frequency-switching mode removes most of the continuum and filter response; a third order polynomial was fit to non-HVC channels outside of -100~km~s$^{-1}$ and +25~km~s$^{-1}$ to remove the remaining residual instrumental baseline.

Some fields  were observed several times to improve the sensitivity and allow  us to estimate  systematic effects brought about by stray radiation,  spectral baselines, or other instrumental problems.  Field centres, total integration times and resultant sensitivities are shown in Table~\ref{fieldsummary}.

\begin{table}[!ht]
\caption{GBT high-latitude observations \label{fieldsummary}}
\smallskip
\begin{center}
{\scriptsize
\begin{tabular}{lrrcrrc}
\noalign{\smallskip}
\tableline
\noalign{\smallskip}
Field name & \multicolumn{2}{c}{Field centre } & Size & $t_{int}$ & Channel & Scan Direction\\
 & $l$ & $b$ & & (sec) & noise (K) & \\
\noalign{\smallskip}
\tableline
\noalign{\smallskip}
Polaris & { 125\fdg00} & { 27\fdg50} & $6\arcdeg\times10\arcdeg$ & 4 & 0.18 & b \\

KnotC & { 138\fdg99} & { 29\fdg42} & $2\arcdeg\times2\arcdeg$ & 8 & 0.12 & b \\

Spider South & { 135\fdg70} & { 29\fdg80} & $12\arcdeg\times8.5\arcdeg$ & 8/4 & 0.11/0.17 & b \\

NEP & { 96\fdg40} & { 30\fdg00} & $12\arcdeg\times12\arcdeg$ & 12 & 0.09 & b \\

KnotB & { 147\fdg64} & {34\fdg77} & $2\arcdeg\times2\arcdeg$ & 8 & 0.14 & b \\

KnotN & {100\fdg28} & { 35\fdg50}& $2\arcdeg\times2\arcdeg$ & 4 & 0.18 & b \\

Ursa Major East & { 155\fdg80} & { 37\fdg00} & $10.5\arcdeg\times6\arcdeg$ & 4 & 0.16 & b \\

Polaris North & { 125\fdg00} & {37\fdg40} & $6\arcdeg\times10\arcdeg$ & 8/4 & 0.12/0.16 & b \\

draco & 92\fdg24  &  38\fdg43 & $5\arcdeg\times5\arcdeg$ & 12 & 0.11 & $\delta$ \\

Ursa Major & { 143\fdg60} & { 38\fdg50} & $9\arcdeg\times9\arcdeg$ & 4 & 0.15 & b \\

KnotA & { 152\fdg62} & { 39\fdg00} & $2\arcdeg\times2\arcdeg$ & 8 & 0.13 & b \\

DRAOdeep & { 134\fdg98} & { 40\fdg00} & $10\arcdeg\times10\arcdeg$ & 
8/4 & 
0.11/0.14 & b \\

N1 & { 85\fdg33} & { 44\fdg28} & $5\arcdeg\times5\arcdeg$ & 8 & 0.13 & b \\

SP & { 132\fdg38} & {47\fdg50}& $5\arcdeg\times5\arcdeg$ & 8 & 0.13 & b \\

g86 & 87\fdg94  &  59\fdg05 & $5\arcdeg\times5\arcdeg$ & 12 & 0.11 & $\delta$ \\

bootesb & 55\fdg98  &  63\fdg87 & $2.5\arcdeg\times4\arcdeg$ & 12 &  0.10 & 
$\delta$ \\

AG & {164\fdg85} & { 65\fdg50} & $5\arcdeg\times5\arcdeg$ & 8 & 0.13 & b \\

bootesd & 56\fdg60  &  65\fdg61 & $2.5\arcdeg\times4\arcdeg$ & 12 & 0.10 & 
$\delta$ \\

Necklace & { 67\fdg82} & { 67\fdg76} & $2\arcdeg\times2\arcdeg$ & 8 & 0.12 & b \\

bootes & 58\fdg11  &  68\fdg57 & $4\arcdeg\times4\arcdeg$ & 
15/6 & 
0.10/0.16 & $\delta$ \\

bootesc & 60\fdg38  &  71\fdg45  & $2.5\arcdeg\times4\arcdeg$ & 12 & 0.10 & 
$\delta$ \\

bootesa & 62\fdg22  &  73\fdg11 & $2.5\arcdeg\times4\arcdeg$ & 8/4 & 
0.16/0.17 & $\delta$ \\

MC & { 56\fdg84} & { -81\fdg50} & $6\arcdeg\times5\arcdeg$ & 8 & 0.14 & b \\
\noalign{\smallskip}
\tableline
\end{tabular}
}
\end{center}
\end{table}

\section{Data Reduction \label{datareduction}}
The GBT is not completely free from ``stray'' 21cm radiation.  Although it lacks the scattering rings that arise from aperture blockage, it still has a foreward spillover lobe \citep{2005LockmanCondon, robishaw_2009} that contains several percent of the telescope's response at 21cm wavelength.
In order to properly correct for the stray radiation, an empirical model of the telescope's sidelobes was developed.  Properties of the main beam were established from electromagnetic calculations, and the sidelobes were measured from observations of the Sun  (Figure~\ref{sunscans}).  
The predicted stray radiation for a given observation, $T_{stray}$, that enters the receiver from directions away from the main beam, is  determined by summing the amount of HI in every direction above the horizon (using the LAB HI survey) weighted by the GBT response in that direction.   Complete information on the procedure is given in  \citet{boothroyd_2010};  here we will summarize some of the results.

\articlefigure[angle=270,scale=0.35]{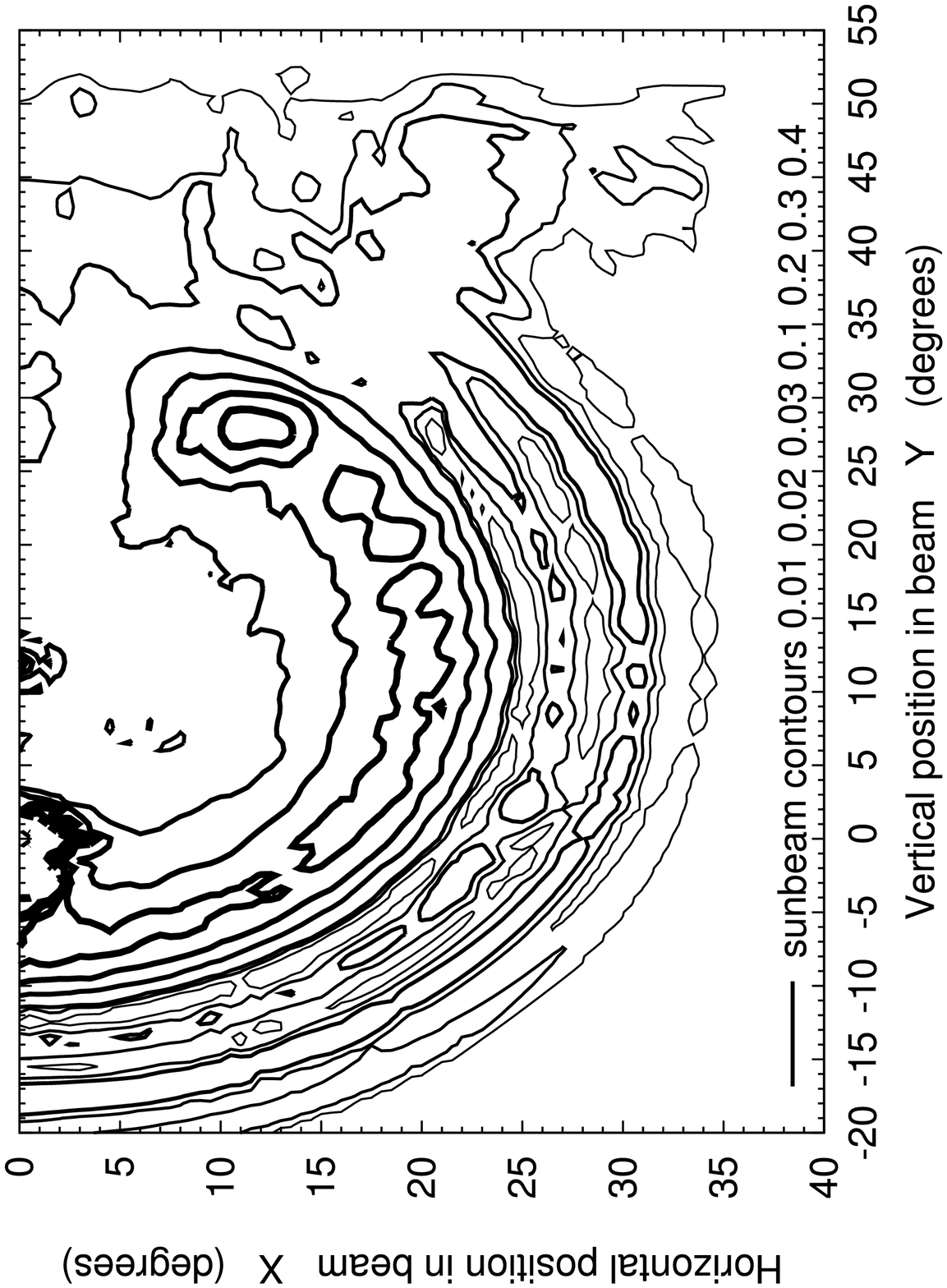}{sunscans}{GBT response pattern from spillover past the subreflector, as measured from scans of the Sun.}

The  antenna temperature scale was established through laboratory calibrations and observations of radio flux calibration sources.  The two methods give results that differ by only  a few percent. 
The GBT antenna temperature, $T_{a}$, is corrected for atmospheric attenuation and converted to brightness temperature, $T_{b}$, adopting a main beam efficiency,  $\eta_{mb}=0.88$, determined from a theoretical calculation of the main beam profile.   The final HI brightness temperature is then 
\begin{equation}
\label{Tb}
T_{b} = \frac{T_{a}^{\star} - T_{stray}}{\eta_{mb}}
\end{equation}
where $T_{a}^{\star}$ is the GBT antenna temperature corrected for atmospheric attenuation, and 
$T_{stray}$ is the calculated sidelobe contribution to the observed antenna temperature profile.

Figure~\ref{prepost} shows the integrated HI from a field as observed over several different sessions with especially large and varying amounts of stray radiation (left), and the same data after the calibration, correction for stray radiation and baseline removal.  The effects of the correction are evident in the reduced number of jumps in the intensity scale.  

\articlefiguretwo{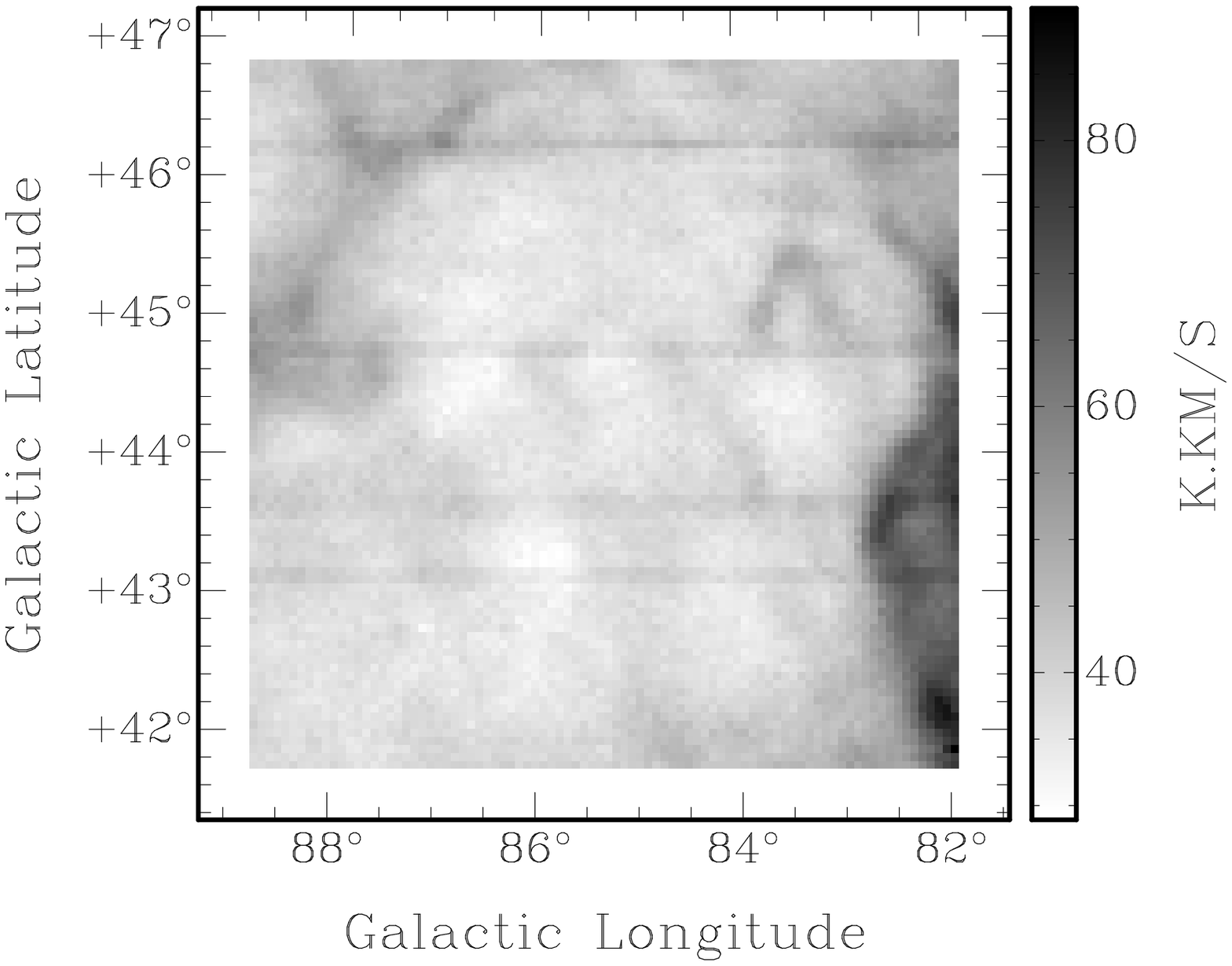}{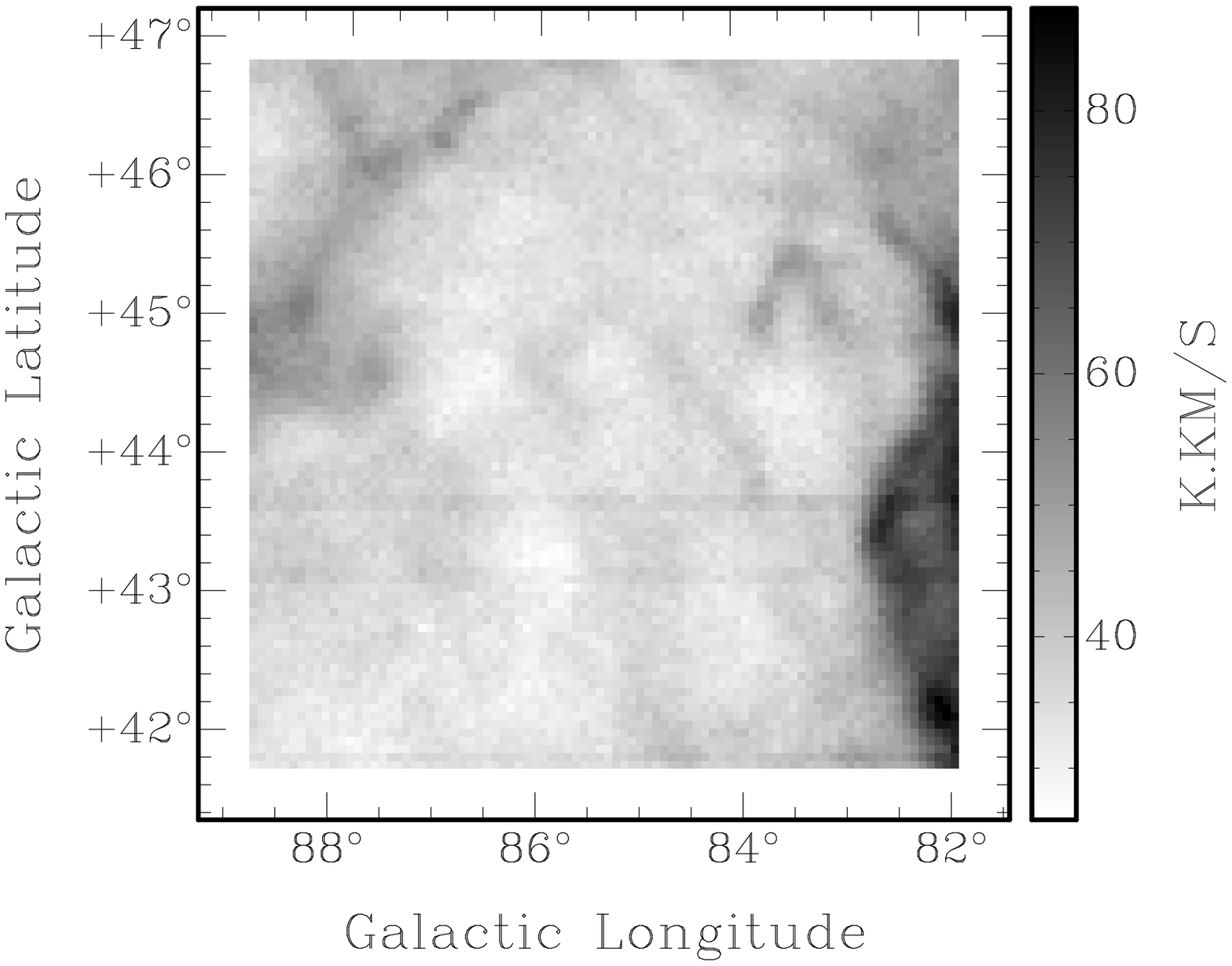}{prepost}{``Worst-case'' low column density integrated GBT cube before stray and baseline removal (left) and after (right), showing a cleaner end product.}

\subsection{General stability \label{stability}}
\citet{robishaw_2009} developed an approximate model for the GBT beam and used this to correct their data for stray radiation.  In so doing, they concluded that the GBT 21cm system was not stable, but had gain fluctuations on the order of 10-15\%, which appeared to vary as a function of LST.
The more detailed model of the stray radiation we have developed allows us to test the gain stability with more confidence, and separate any  true gain variability from other effects.
The telescope, receiver, and IF systems were identical to those used by \citet{robishaw_2009}, but we used the GBT Spectrometer rather than the Spectral Processor as the detector.

Over the course of the high Galactic latitude observations, which occured in many sessions over a three-year period, two fields were revisited regularly to test the consistency of the calibration: S6 and S8.   
After reducing these data using our correction for stray radiation, we
find the measurements to be consistent to within a few percent,  with no evidence of a correlation between HI brightness temperature and date or LST. The scatter in the measurements is entirely consistent with our estimates for the uncertainties arising from noise, spectral baselines, and residual stray radiation.  We do not reproduce, nor can we understand, the ``gain variations'' reported by \citet{robishaw_2009}.  We also note that many tens of thousands of HI spectra have been measured with the GBT by a number of  observers, using both the Spectral Processor and Spectrometer, without detecting significant gain variations (e.g., \citet{2005LockmanCondon, 2008Chynoweth}).

\subsection{A quick calibration check}

To confirm the accuracy of the final conversion to brightness temperature (as in Equation~\ref{Tb}), we compare our mean integral of S6 and S8 with those of \citet{williams_1973} and \citet{kalberla_1982}: $\sum_{i}T_{b}^{i} \Delta v=299~$K~km~s$^{-1}$, and $846$~K~km~s$^{-1}$, respectively.
The mean integrals of our GBT data are 289~K~km~s$^{-1}$ for S6, and 845~K~km~s$^{-1}$ for S8 in remarkable agreement with both \citet{williams_1973} and \citet{kalberla_1982}.  

\articlefigure[scale=0.5]{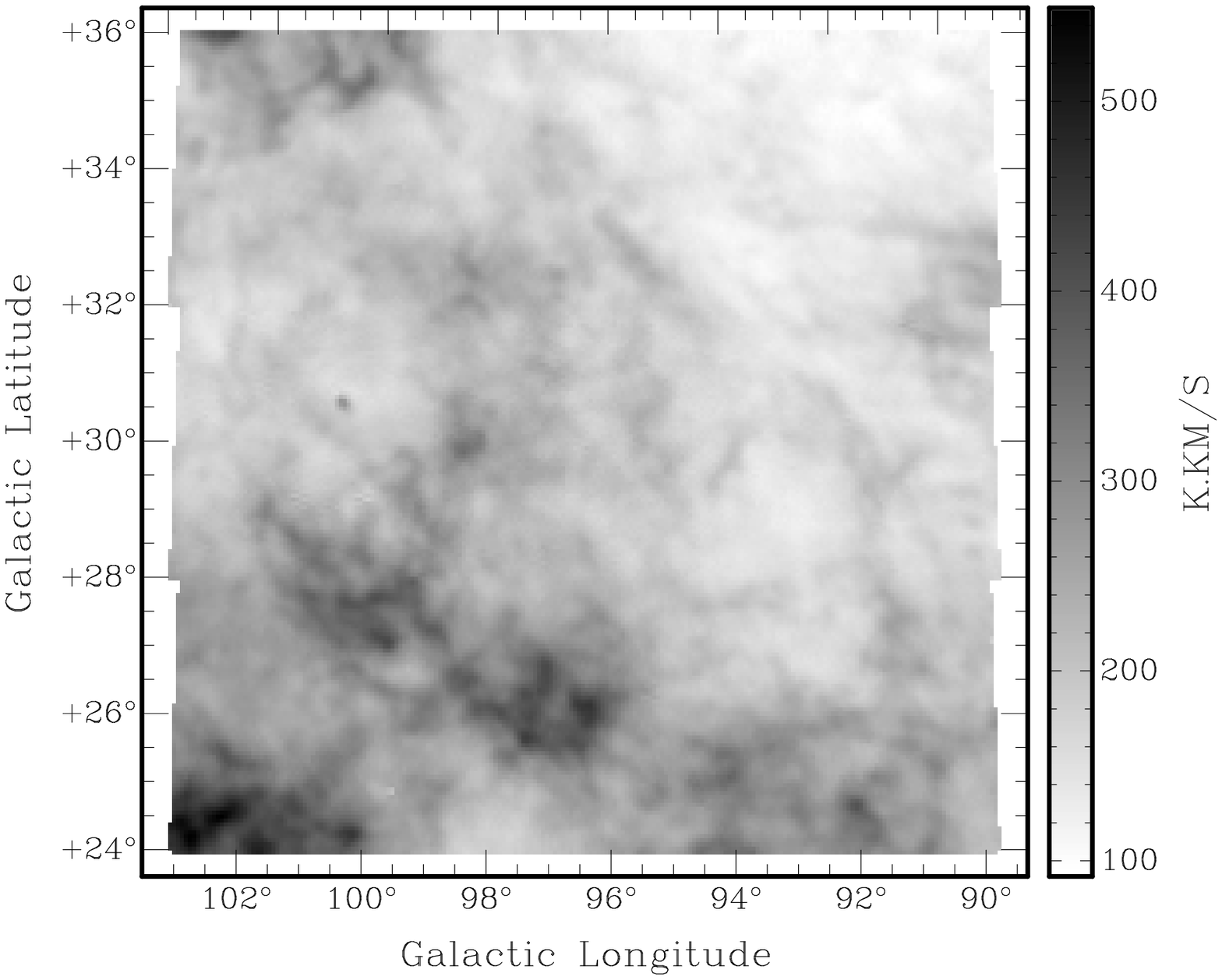}{nep}{Total integrated HI intensity (zeroth moment) map of NEP cube}

\section{Results}
\subsection{Data quality}
The  integrated HI emission over the 12\deg by 12\deg\  NEP mosaic is shown in Figure~\ref{nep}.  It is clear from this mosaic of observations at different LSTs (and thus very different positioning of the beam's sidelobes on the sky) that residual effects, which would appear as stripes or discordant spots, are small.

\subsection{Dust-gas correlation}
In the optically thin case the integrated line intensity, $W_{HI}$ (in K~km~s$^{-1}$), can be converted to a column density, $N_{HI}$:
\begin{equation}
N_{HI} = 0.01823 W_{HI}
\end{equation}
where $N_{HI}$ is in units of $10^{20}$ cm$^{-2}$.

\articlefigure[scale=0.4]{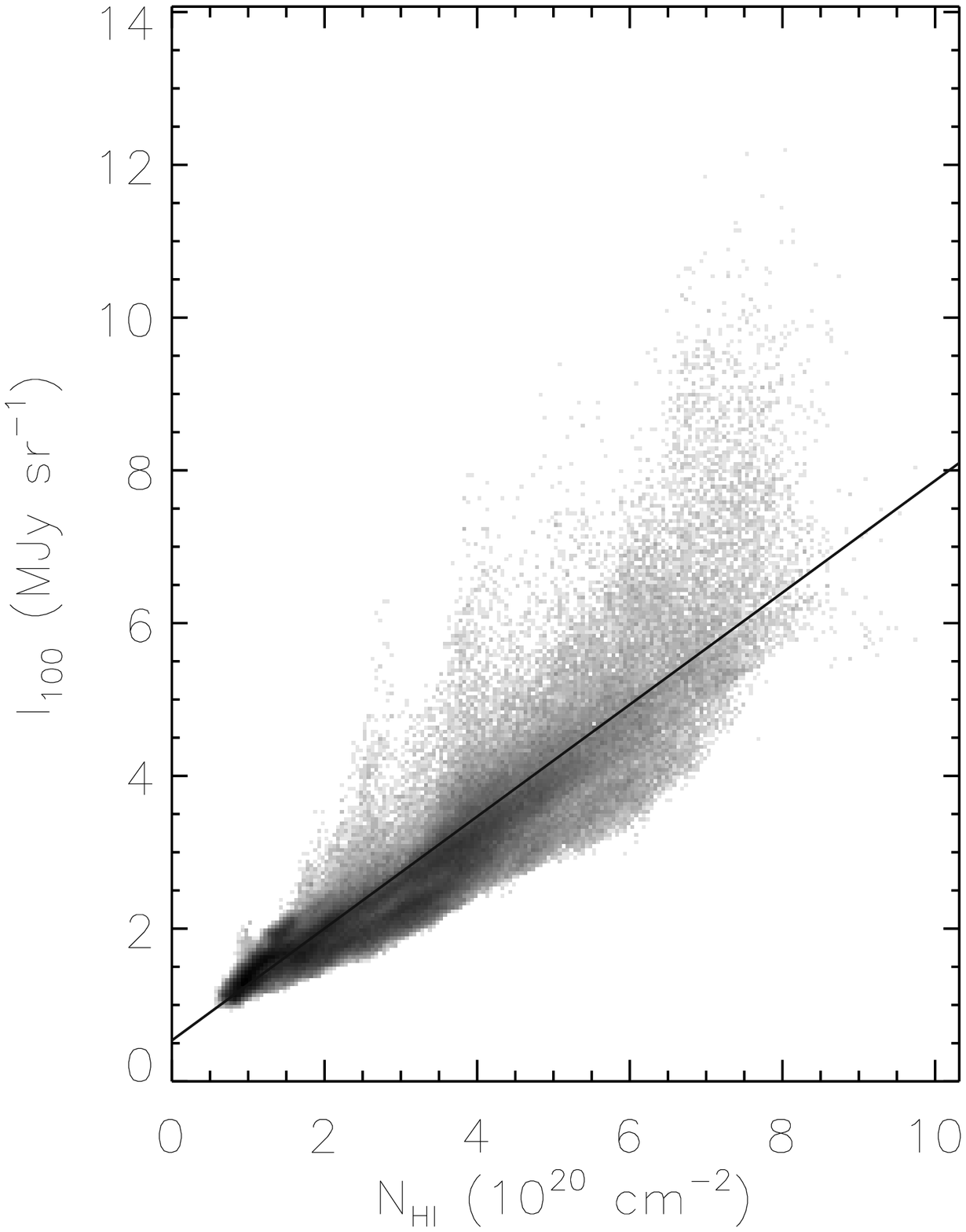}{dustgas}{GBT $N_{H I}$ correlation with IRIS 100$\mu$m over more than 800 square degrees of high Galactic latitude sky}

The GBT HI/IRIS 100~$\mu$m correlation is shown in Figure~\ref{dustgas} for every independent observation, covering over 800 square degrees of high Galactic latitude sky.
The correlation is quite strong and the slope ($0.67$ MJy sr$^{-1}$) is consistent with that found by \citet{boulanger_1996} ($0.69\pm0.03$).  However, there are clear deviations from the average.   At high column densities, an infrared excess can be attributed to the presence of molecular hydrogen \citep{boulanger_1988,joncas_1992,boulanger_1996}, but this does not explain the scatter seen throughout Figure~\ref{dustgas}.  Does the scatter arise from real emissivity variations across the sky,  or possibly emissivity variations along the line-of-sight?    The former analysis is covered in \citet{martin_2010} while the latter is discussed in the light of velocity tomography in the following section.

\subsection{Velocity tomography}
A number of the GBT fields were selected for the presence of IVC or HVC gas.  If these clouds can be separated from the gas consistent with Galactic rotation (the 'local' gas) and from each other, a series of dust-emissivity-related coefficients ($a_{\lambda}$, $b_{\lambda}$, ...) can be determined for each IR or sub-mm map, $I_{\lambda}$.  An example of such a velocity separation is shown in Figure~\ref{nepvelocities}.
The coefficients are then derived from solutions to
\begin{equation}
\label{emissivities}
I_{\lambda} = a_{\lambda} + b_{\lambda} N_{HI}^{b} + c_{\lambda} N_{HI}^{c} + ... ,
\end{equation}
where component $a$ is the cosmic infrared background.

\articlefigurethree{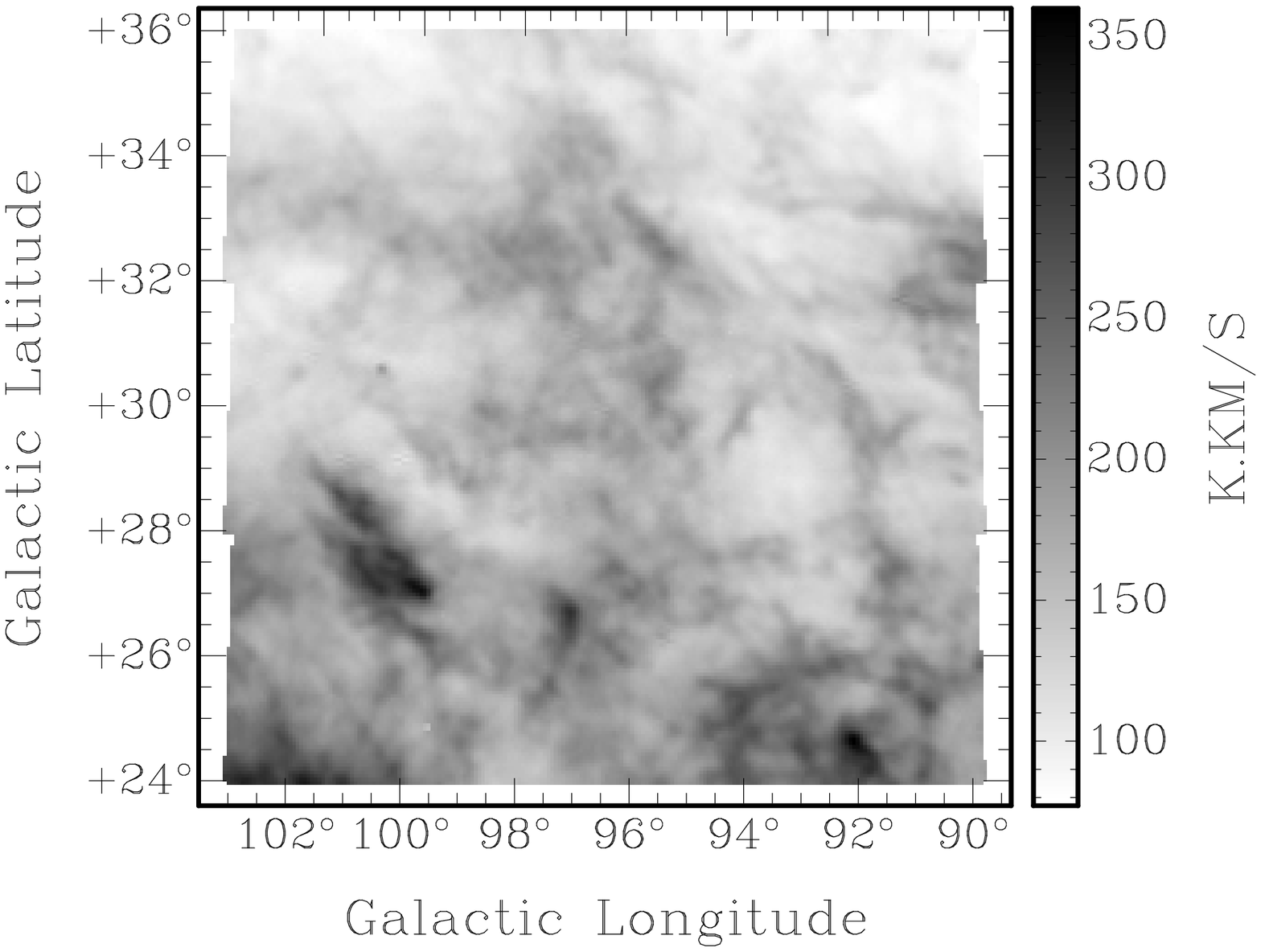}{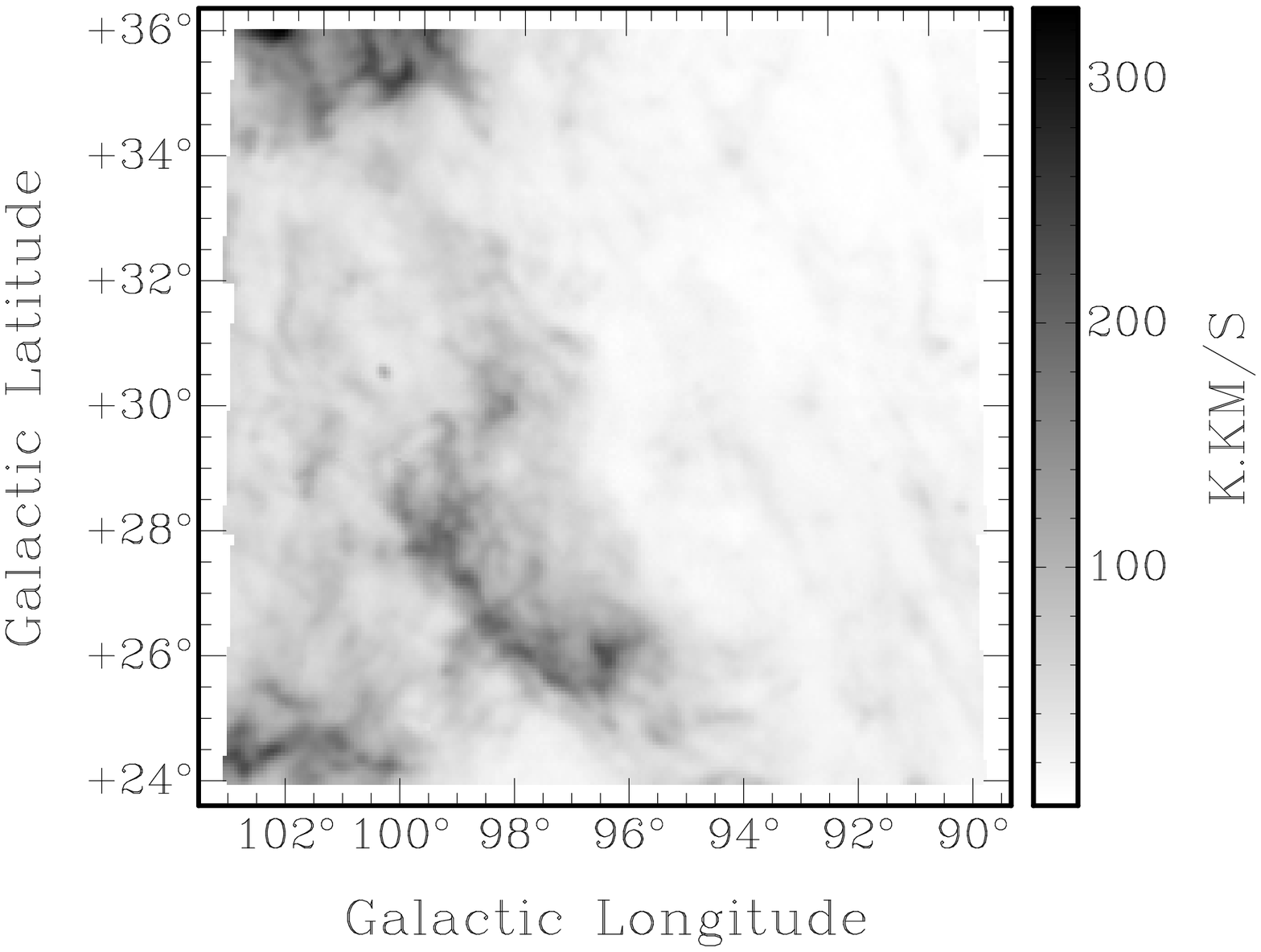}{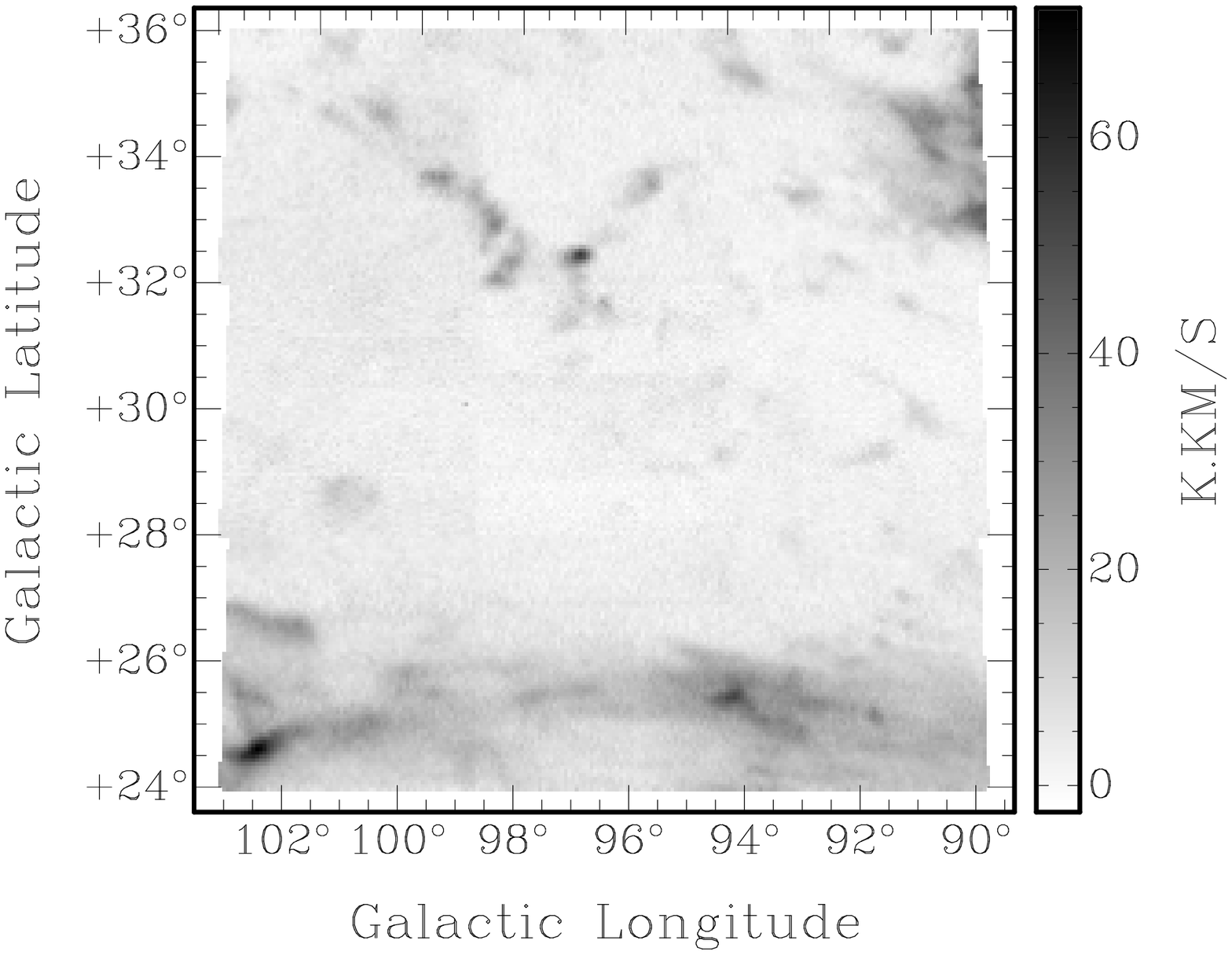}{nepvelocities}{Local (left), IVC (centre) and HVC (right) components of the 21cm emission in the NEP field.  Note differences in intensity scale.}

With knowledge of these emissivities for each given component ($a$,$b$,$c$,...) at a variety of wavelengths, a spectral energy distribution (SED) can be fit and dust properties for the given component can be determined.  With the anticipated contribution from Planck in the sub-mm, this is on the verge of becoming an even more useful analysis of the dust content of, and dust evolution in our Galaxy.  For now, this analysis has been done using IRIS (reprocessed IRAS) data \citep{mamd_2005} at 60~$\mu$m and 100~$\mu$m.
The sample field (NEP, as in Figure~\ref{nepvelocities}) shows evidence of enhanced 60~$\mu$m/100$\mu$m colour in the IVC as compared to the local gas as has been seen in the Draco IVC \citep{herbstmeier_1993}.  This is an indicator of dust processing with a slight increase in the relative contribution from small grains to the SED.
In another field, Bo\"{o}tes, the IVC dust not only appears to be processed, but it appears to be exposed to a different (stronger) external radiation field.  Understanding this in the context of the assumed constant interstellar radiation field at high Galactic latitude is key.  Apparently, not all IVC gas is necessarily of the same origin, or at the least it cannot all be described as having an identical mixture of gas and dust, heated identically.

\section{The future}
In addition to studying the correlation between gas and dust, these GBT HI maps offer a large area over which to study the gas turbulence, the interaction between phases of the ISM and the properties of compact HVCs, amongst other things.  This area of study promises to be quite fruitful over the next few years.

\acknowledgements 
The National Radio Astronomy Observatory is operated by Associated Universities, Inc., under a cooperative agreement with the National Science Foundation.

\bibliography{ms}

\end{document}